\documentstyle[11pt,aaspp4,flushrt]{aastex}

\slugcomment{\it Astrophysical Journal 755, 173, 2012 Aug. 20}
\shortauthors{Eisenhardt et al.}
\shorttitle{The First Hyper-Luminous Infrared Galaxy Discovered by WISE}

\def\longw1814{WISE\,J181417.29+341224.9}
\def\w1814{WISE 1814+3412}
\def\deg{\ifmmode {^{\circ}}\else {$^\circ$}\fi}
\def\spose#1{\hbox to 0pt{#1\hss}}
\def\simlt{\mathrel{\spose{\lower 3pt\hbox{$\mathchar"218$}}
     \raise 2.0pt\hbox{$\mathchar"13C$}}}
\def\simgt{\mathrel{\spose{\lower 3pt\hbox{$\mathchar"218$}}
     \raise 2.0pt\hbox{$\mathchar"13E$}}}

\newcommand{\spitzer}{{\it Spitzer}}

\def\plotfiddle#1#2#3#4#5#6#7{\centering \leavevmode
\vbox to#2{\rule{0pt}{#2}}
\includegraphics{#1}}

\begin{document}

\title{The First Hyper-Luminous Infrared Galaxy Discovered by WISE}

\author{
Peter~R.~M.~Eisenhardt\altaffilmark{1},
Jingwen~Wu\altaffilmark{1,2},
Chao-Wei~Tsai\altaffilmark{3},
Roberto~Assef\altaffilmark{1,2},
Dominic~Benford\altaffilmark{4},
Andrew~Blain\altaffilmark{5},
Carrie~Bridge\altaffilmark{6},
J.~J.~Condon\altaffilmark{7},
Michael~C.~Cushing\altaffilmark{8},
Roc~Cutri\altaffilmark{3},
Neal~J.~Evans~II\altaffilmark{9},
Chris~Gelino\altaffilmark{3},
Roger~L.~Griffith\altaffilmark{3},
Carl~J.~Grillmair\altaffilmark{3},
Tom Jarrett\altaffilmark{3},
Carol~J.~Lonsdale\altaffilmark{7},
Frank~J.~Masci\altaffilmark{3},
Brian~S.~Mason\altaffilmark{7},
Sara~Petty\altaffilmark{10},
Jack~Sayers\altaffilmark{6},
S.~Adam~Stanford\altaffilmark{11},
Daniel~Stern\altaffilmark{1},
Edward~L.~Wright\altaffilmark{10},
Lin~Yan\altaffilmark{3}
}

\altaffiltext{1}{Jet Propulsion Laboratory, California Institute
of Technology, MS 169-327, 4800 Oak Grove Drive, Pasadena, CA 91109
[e-mail: {\tt Peter.R.Eisenhardt@jpl.nasa.gov}]}

\altaffiltext{2}{NASA Postdoctoral Program Fellow}

\altaffiltext{3}{Infrared Processing and Analysis Center, California
Institute of Technology, Pasadena, CA 91125}

\altaffiltext{4}{NASA Goddard Space Flight Center, Greenbelt, MD 20771}

\altaffiltext{5}{Department of Physics \& Astronomy, University of Leicester, Leicester, LE1 7RH, United Kingdom}

\altaffiltext{6}{Division of Physics, Math and Astronomy, California
Institute of Technology, Pasadena, CA 91125}

\altaffiltext{7}{National Radio Astronomy Observatory, 520 Edgemont Road, Charlottesville, VA 22903}

\altaffiltext{8}{University of Toledo, Toledo, OH 43606}

\altaffiltext{9}{Department of Astronomy, University of Texas, Austin, TX 78712}

\altaffiltext{10}{University of California, Los Angeles, CA 90095}

\altaffiltext{11}{University of California, Davis, CA 95616}

\begin{abstract}

We report the discovery by the Wide-field Infrared Survey Explorer 
of the $z=2.452$ source \longw1814,   
the first hyperluminous source found in the WISE survey. 
\w1814\ is also the prototype for an all-sky sample of 
$\sim1000$ extremely luminous ``W1W2-dropouts" 
(sources faint or undetected by WISE at 3.4 and $4.6\ \mu$m 
and well detected at 12 or $22\ \mu$m).  
The WISE data and a $350\ \mu$m detection give a 
minimum bolometric luminosity of $3.7\times10^{13}L_\odot$,
with $\sim10^{14}L_\odot$ plausible.
Followup images reveal four nearby sources:
a QSO and two Lyman Break Galaxies (LBGs) at $z=2.45$, and an M dwarf star.  
The brighter LBG dominates the bolometric emission.  
Gravitational lensing is unlikely given
the source locations and their different spectra and colors.    
The dominant LBG spectrum
indicates a star formation rate $\sim300 M_\odot {\rm yr}^{-1}$, 
accounting for $\simlt10\%$ of the bolometric luminosity. 
Strong $22\ \mu$m emission relative to $350\ \mu$m 
implies that warm dust contributes significantly to the luminosity, while 
cooler dust normally associated with starbursts 
is constrained by an upper limit at 1.1 mm.  
Radio emission is $\sim10\times$ above the far-infrared/radio correlation, 
indicating an active galactic nucleus is present.    
An obscured AGN combined with 
starburst and evolved stellar components 
can account for the observations. 
If the black hole mass follows the local $M_{\rm BH}$-bulge mass 
relation, the implied Eddington ratio is $\simgt4$.  
\w1814\ may be a heavily obscured object 
where the peak AGN activity occurred prior to the peak era of star formation.  

\end{abstract}


\section{Introduction
\label{sec:intro}}

The Wide-field Infrared Survey Explorer (WISE) launched on 2009 Dec. 14, 
and began surveying the sky on 2010 Jan. 14, completing its first full coverage 
in July 2010.  WISE achieves $5 \sigma$ point source sensitivities of 
better than 0.08, 0.11, 1, and 6 mJy at 3.4, 4.6, 12 and $22\ \mu$m  
(hereafter referred to as W1, W2, W3, and W4) in a single coverage on the ecliptic, 
consisting of 8 or more exposures at each sky location \citep{Wright:10}.  
Sensitivity improves away from the ecliptic 
due to denser exposure overlap and lower zodiacal background.
The survey continued in W1 and W2 after the cryogen was exhausted 
at the end of Sept. 2010, and concluded 2011 Feb. 1, having achieved 
two complete sky coverages in these two bands.  The 
WISE all-sky data release was issued on 2012 Mar. 
14\footnote{\tt http://wise2.ipac.caltech.edu/docs/release/allsky}. 

The primary science objectives for WISE are to identify 
the coldest and nearest brown dwarfs to the Sun
\citep[see e.g., ][]{Mainzer:11a, Cushing:11, Kirkpatrick:11}, 
and the most luminous, dusty, forming galaxies 
(Ultra-luminous Infrared Galaxies or ULIRGs). 
With regard to the most luminous objects, 
the dominant sources of energy production in the Universe 
are fusion in stars and 
gravitational accretion onto super-massive black holes.  
The tight correlation between the masses of  
super-massive black holes at the centers of galaxies, 
and the masses of the stellar bulges in these galaxies
\citep{Magorrian:98, Ferrarese:00,Gebhardt:00}, 
implies the two formation processes are intimately connected.   
Bulge stellar populations are old and quiescent today, 
and so must have formed in the distant past, 
when the cosmic star formation rate was 
much higher \citep[e.g.,][]{Hopkins:04}.  
Similarly the peak era of accretion 
onto super-massive black holes was at redshifts $z\sim2$, 
when luminous quasars were common \citep[e.g.,][]{Richards:06a, Assef:11}. 
There is substantial evidence that dust absorbs 
much of the UV/optical luminosity generated by 
the formation of massive galaxies and their central black holes at $z > 1$ 
\citep[e.g.,][]{Blain:04, LeFloch:05, Stern:05b, Hickox:07}.  
The dust is heated in the process, and most of the luminosity 
emerges at IR wavelengths, creating a ULIRG. 

The most extreme examples of ULIRGs are therefore 
likely associated with the major formation events 
of the most massive galaxies. Such objects should appear in 
the sensitive infrared all-sky WISE survey, and may be missed
in surveys with substantially smaller areas by e.g., \spitzer\ 
and {\it Herschel}.     
Here we report on \longw1814\ (hereafter WISE 1814+3412),
the first hyper-luminous infrared galaxy ($L_{\rm IR} > 10^{13} L_\odot$) 
discovered by WISE. 
 
Magnitudes are converted to flux densities using zeropoint values of
3631 Jy for the AB system {\it g'} and {\it r'} bands.  Other magnitudes are on
the Vega system, using zeropoints of 1594 and 666.7 Jy for the 
2MASS system {\it J} and ${\it K_s}$ bands; 280.9 and 179.7 Jy for {\it Spitzer} IRAC
[3.6] and [4.5]; and 306.7, 170.7, 29.04 \& 8.284 Jy for
WISE W1 through W4 respectively \citep{Wright:10}. 
Luminosities are calculated assuming 
$\Omega_M = 0.3$, $\Omega_\Lambda = 0.7$, and
$H_0 = 70\ {\rm km\ s}^{-1} {\rm Mpc}^{-1}$.

\section{Selection Criteria and Followup Observations\label{sec:wise}}

\begin{figure}
\plotfiddle{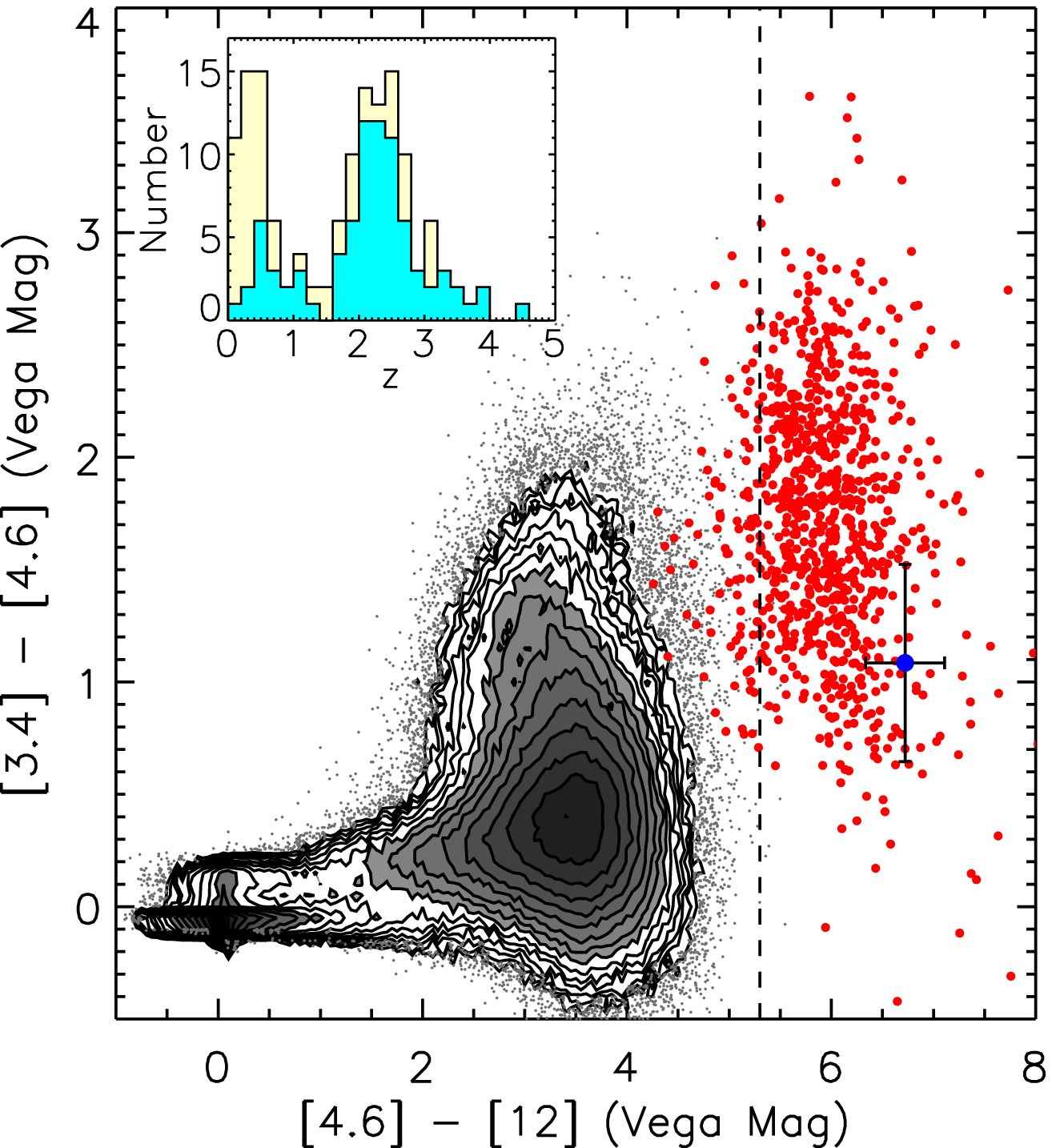}{5.0in}{0}{100}{100}{-200}{-10}
\caption{WISE color-color plot of the 226,017 sources with $< 0.3$ mag errors 
in W1 (3.4 $\mu$m), W2 (4.6 $\mu$m), and W3 (12 $\mu$m), in the 313.4 square degree area 
of the all-sky release catalog with Galactic latitude ${\rm b} > 80\deg$ (black points and gray regions).
These represent about 8\% of all WISE catalog sources in this region.
The larger red points are for 907 W1W2-dropouts selected over the whole sky,
excluding the Galactic plane and bulge, an area $\sim100$ times larger, 
illustrating the rarity of the W12drop population. 
The inset shows the histogram of available redshifts for the W12drop sample, with
the cyan shading corresponding to W1W2-dropouts not seen in the optical digitized sky survey.
The contours are in surface density at power law steps of $\sqrt{2}$,
with the outermost contour at 10 sources per $0.05\ {\rm mag} \times 0.05$ mag. 
The colors plotted for the W1W2-dropouts use {\it Spitzer} 3.6 and 4.5 $\mu$m 
followup data since they are generally not detected at 3.4 and 4.6 $\mu$m by WISE. 
{\it Spitzer} data were converted to the WISE system using 
$[3.4] - [4.6] = W1 - W2 = 1.4([3.6] - [4.5])$, and
$[4.6] - [12] = W2 - W3 = 0.07 + [4.5] - [12]$. 
The large blue symbol with error bars is for \w1814.  
The vertical dashed line at $W2-W3=5.3$ is one of the selection criteria for W1W2-dropouts; 
some are bluer than this because they satisfied the $W2-W4 > 8.2$ criterion. 
\label{fig:colcol}}
\end{figure}

Early searches for the most luminous galaxies with WISE data
included the investigation of outlier populations, 
among them objects which were only well detected in W3 and W4, including \w1814.
This approach proved highly successful (Figure~\ref{fig:colcol}), leading to a variety of followup
programs now underway on sources which are much fainter in W1 (3.4 $\mu$m) 
and W2 (4.6 $\mu$m) than W3 (12 $\mu$m) or W4 (22 $\mu$m).
The selection criteria for these ``W1W2-dropouts" are 
W1 $> 17.4\ (<34\ \mu$Jy), and either:
a) W4 $<7.7\ (>6.9$ mJy) and ${\rm W2} - {\rm W4} > 8.2$; 
or
b) W3 $< 10.6\ (>1.7$ mJy) and ${\rm W2} - {\rm W3} > 5.3$.
W1W2-dropouts must also have at least 7 individual WISE exposures available for measurement 
in W3 or W4, be more than  
$30\deg $ from the Galactic center and $10\deg $ from the Galactic plane,
be free of artifacts flagged by the WISE pipeline, not be associated with
known asteroids \citep[including asteroids discovered by WISE;][]{Mainzer:11b},
and finally pass a series of visual inspections of both the individual exposures 
and coadded images for evidence of spurious sources.  
Figure~\ref{fig:colcol} illustrates the rarity of W1W2-dropouts in WISE color-color space.
The sources have a surface density $\simlt 0.03\ {\rm deg}^{-2}$, or $<10^{-5}$ that of WISE sources in general.
Figure~\ref{fig:colcol} also shows the distribution of the 143 redshifts currently known for
the sample. Over 60\% of W1W2-dropouts are at $z > 1.6$, with 
most of the low redshift population optically bright, as noted by \citet{Bridge:12}.

\w1814\ satisfies all these selection criteria, 
and was initially identified in 2010 May using similar criteria and
available data.
Of the 37 sources which satisfied the criteria at the time, 
\w1814\ 
was the brightest in W4.
Figure~\ref{fig:w12d} shows cutouts of images in the four WISE bands of \w1814, 
and Table~\ref{table:phot} gives W3 and W4 PSF fitted photometry for the source.
In the two shorter bands the $2\sigma$ limit for W1 is formally 
$> 18.67\ (<11\ \mu$Jy) 
although magnitudes at these levels appear to underestimate the true flux density, 
while there is a marginal detection at $\rm{W2}=17.17 \pm 0.49\ (23 \pm 11\ \mu$Jy).

\subsection{Images at Other Wavelengths
\label{sec:OtherData}}

Followup imaging observations of \w1814\ have been carried out from optical to radio wavelengths, 
with results summarized in Table~\ref{table:phot}.  Details of these observations follow.

\subsubsection{Optical Imaging \label{sec:optical}}

Optical imaging of \w1814\ 
using {\it g'} and {\it r'} filters
was obtained 
on UT 2011 May 28 with the Mosaic-1.1
Imager \citep{Sawyer:10} on the KPNO 3.8~m diameter\footnote{Although
generally referred to as the KPNO 4m telescope, 
the clear aperture diameter of the
Mayall telescope is 3.797m, as documented at 
{\tt http://www-kpno.kpno.noao.edu/kpno-misc/mayall\_params.html}}
Mayall telescope. 
The pixel size was $0\farcs26$ at the 
center of the CCD mosaic, declining to $0\farcs245$ at the edge of the 36' field. 
Three dithered 500~s exposures in the two filters 
were obtained in clear conditions with sub-arcsecond seeing.
Pipeline processed stacked images resampled onto uniform
$0\farcs25$ pixels 
were retrieved from the NOAO Science Archive, 
yielding a FWHM image size of 
$0\farcs75$ in {\it g'} and $0\farcs70$ in {\it r'}.  
Small ($\sim0\farcs2$) astrometric offsets were made to the image world coordinate system (WCS) to
remove any differences with 2MASS catalog positions of objects in the field.
Because the \w1814\ field is not included in the SDSS,
flux calibration was obtained using several short (30~s) observations 
in the two filters of another W12drop with SDSS coverage 
$10\deg$ away.  These were
immediately followed by 30~s exposures of \w1814\ in {\it g'},
then by full depth 500~s exposures of \w1814\ in {\it r'}, and finally
by full depth 500~s exposures of \w1814\ in {\it g'}.  All observations
of the two W12drop fields were at an airmass of 1.04 or less.  
No significant change in zero point was detected 
between the 30~s and 500~s {\it g'} exposures on \w1814.

The lower left panel of Figure~\ref{fig:w12d} 
shows a $2\arcmin\times2$\arcmin\ region 
of the reduced {\it r'} image centered on \w1814,
and the upper left panel of Figure~\ref{fig:nirc2} shows a
zoomed in view of the {\it g'} image covering $20\arcsec \times20$\arcsec.
Several objects are apparent near the \w1814\ coordinates,
with the closest being a somewhat extended source 
(labeled ``A" in the upper left panel of Figure~\ref{fig:nirc2}),
two unresolved sources (``B" and ``C") to the north, a fainter
source (``D") a few arcsec to the southwest, and two other faint sources
farther to the southwest and east.  These sources are also seen in
the {\it r'} image.
Components A, B, C, and D were also seen in optical images 
obtained on UT 2010 June 14 at Palomar Observatory 
using the Hale 5.08~m telescope with the Large Format Camera \citep{Simcoe:00} at prime focus.  
The Palomar images were used to position spectroscopic slits
as described in \S~\ref{sec:spectrum}, but because of their relatively
poor image quality and non-photometric conditions, are not further discussed here.

Because  of the close proximity of source B to A and the fact that source A is extended,
the optical photometry for source A reported in Table~1 
is from a $7\farcs5\ $ diameter aperture
in an image in which sources B and C have been PSF subtracted.
The optical photometry for source B in Table~1 is from PSF fitting.

\subsubsection{Near-IR Imaging \label{sec:nearIR}}

\w1814\ was targeted for followup in $K_s$ with the NIRC2 camera 
on the Keck~II telescope 
using laser guide star adaptive optics \citep{Wizinowich:06} on UT 2010 July 1,
and in {\it J} on UT 2011 July 20.  
A total of 30 minutes of integration were obtained in $K_s$ using dithered 1 minute exposures,
and 18 minutes in {\it J} using 2 minute exposures, 
with a pixel scale of $0\farcs0397$ giving a 40\arcsec\ field.   
The $K_s$ data were obtained under photometric 
conditions and $0\farcs35$ seeing, while for {\it J} the seeing was $>1$\arcsec\ and 
the sky was cloudy at the end of the  night.  An $R=15.5$ star $33\arcsec$
from \w1814\ was used to provide tip-tilt correction.  

Each individual image was sky subtracted using a median 
of the 4 preceding and 4 following images.
The sky-subtracted images were roughly registered using 
the recorded position of the AO tip-tilt sensor stage,
and then fine tuned by minimizing their residuals relative to the first image. 
The final mosaic was formed from the median of aligned images.    
Photometric calibration was achieved using several well exposed stars 
in common with the wider field Palomar/WIRC data described below,
but because these exhibited an rms dispersion of 0.2 mag, that dispersion was
imposed as a minimum photometric error.
Astrometric calibration was achieved using sources in common with
the optical {\it r'} image which had been registered to the 2MASS
frame (\S~\ref{sec:optical}).   
The upper right panel of Figure~\ref{fig:nirc2} shows the central
$20\arcsec \times20$\arcsec\ of the resulting $K_s$ image. 
Components A, B, C, and D are detected in
$K_s$, while in {\it J}, component A is faint and D is undetected.
In both images component A is clearly extended relative to components
B and C.
Photometry was measured in a 50 pixel ($1\farcs99$) diameter aperture
for component A and a 20 pixel ($0\farcs79$) diameter aperture for component B,
centered on the peak of each component.  

A $K_s$ image covering $8\farcm7$ on the \w1814 field was obtained
on UT 2010 July 27 and in {\it J} on 2011 Sept. 17 
with the Wide-field Infrared Camera \citep[WIRC;][]{Wilson:03}
at Palomar Observatory using the Hale 5.08~m telescope.  For both images the seeing was
sub-arcsecond and the pixel size was $0\farcs25$. The total integration
time was 18 minutes for $K_s$ and 5 minutes for {\it J},  
with 1 minute per dither.  The data were flatfielded and
coregistered, and calibrated to 2MASS.  Although \w1814 was only
faintly detected in $K_s$ and not at all in {\it J} in these data, 
the WIRC images enabled the photometric calibration of the deeper
NIRC2 images to be tied to 2MASS.

\subsubsection{Spitzer \label{sec:Spitzer}}

\w1814\ was observed by \spitzer\ at [3.6] 
and [4.5] on 2010 July 22 using Director's Discretionary Time (Program \# 549).  
The source was observed with the $1\farcs2$ pixels and 5\arcmin\ field
of each IRAC band using twelve exposures each 100~s long, 
employing the medium scale Reuleaux dither pattern.
Figure~\ref{fig:w12d} shows the \spitzer\ pipeline post-BCD processed image
of \w1814 at $3.6\ \mu$m, which is resampled onto $0\farcs6$ pixels,
while Figure~\ref{fig:nirc2} shows the $4.5\ \mu$m image. 

The multiple components seen in 
Figure~\ref{fig:nirc2} are present in the \spitzer\ data,
with the relative prominence of component A increasing at longer wavelengths.
Similar to the optical imaging (\S~\ref{sec:optical}),
the \spitzer\ photometry for source A reported in Table~1
is from a 12\arcsec\ diameter aperture
in an image in which sources B and C were PSF subtracted, 
while the photometry for source B is from PSF fitting.
The [4.5] detection is consistent with the WISE W2 given above,
while the [3.6] detection implies that the W1 $2\sigma$ limit
underestimates the true flux density.
A minimum photometric error of 0.1 mag has been imposed on the
\spitzer\ photometry.

\subsubsection{Far-Infrared Data \label{sec:IRAS}}

\w1814\ is undetected by IRAS. 
Using SCANPI\footnote{\tt http://irsa.ipac.caltech.edu/IRASdocs/scanpi}, 
the $2\sigma$ upper limits are
$< 40$~mJy at 12 and $25\ \mu$m, $<100$~mJy at $60\ \mu$m, and $<600$~mJy at $100\ \mu$m.
These limits are consistent with the W3 and W4 detections of \w1814.

\subsubsection{Caltech Submillimeter Observatory Data \label{sec:CSO}}

Continuum observations of \w1814\ at $350\ \mu$m 
were obtained on 2010 July 13 and 23,  and at $450\ \mu$m on 2010 Sept. 12 and 13, 
using the Submillimeter High Angular Resolution Camera II (SHARC-II) 
at the 10.4 m telescope of the Caltech Submillimeter Observatory (CSO) 
at Mauna Kea, Hawaii \citep{Dowell:03}. 
SHARC-II uses a bolometer array with 32 $\times$ 12 pixels, 
resulting in a 2.59\arcmin\ $\times$ 0.97\arcmin\ field of view. 
The beam sizes of SHARC-II at 350 and 450 $\mu$m 
are 8\farcs5 and 10\farcs9 respectively.

Atmospheric transmission at 350 $\mu$m and 450 $\mu$m is poor when
the opacity at 225 GHz ($\tau_{225}$) is $> 0.07$.
\w1814\ was observed near transit for 160 minutes at 350 $\mu$m with $\tau_{225} = 0.035 - 0.06$  
and for 90 minutes at 450 $\mu$m with $\tau_{225} = 0.045$.
The telescope was scanned in a Lissajous pattern that keeps the source within the FOV. 
The Dish Surface Optimization System \citep[DSOS;][]{Leong:06} 
was used to correct the dish surface figure for 
imperfections and gravitational deformations 
as the dish moves in elevation during observations. 

The SHARC-II data were reduced using 
version 2.01-4 of the Comprehensive Reduction Utility for SHARC-II \citep[CRUSH;][]{Kovacs:06a}. 
Pointing was checked every 1-2 hours and corrected to $<2\farcs0$ in altitude
and $<2\farcs5$ in azimuth   
using Uranus when available, 
and with a secondary calibrator, 
K3-50 (a massive star forming region close to \w1814).  
The processed $350\ \mu$m image is
shown in Figure~\ref{fig:w12d}.  Flux calibration was obtained using Uranus.
We derive a flux density at 350 $\mu$m of $33 \pm 9$ mJy for \w1814.
At 450$\ \mu$m the detection was not significant, with a flux density of $15\pm10$ mJy.
We report in Table~1 a 95\% confidence upper limit of 35~mJy at $450\ \mu$m, 
based on the statistical methods described by \citet{Feldman:98}.

\w1814\ was observed at 1.1~mm with the CSO using 
the Bolocam instrument for a total of ten hours
on UT 2010 June 18 and 19.
Bolocam is a large format camera with 144 detectors, an $8 \arcmin$ FOV,
and a $30\arcsec$ FWHM \citep{Haig:04}.
The telescope was scanned in a Lissajous pattern 
keeping the source in the FOV.
Pointing was checked and corrected to 6\arcsec\ using
frequent observations of the nearby bright objects K3-50
and 3C345 following methods described in \citet{Sayers:09}.
Flux calibration, estimated to be accurate to $\simeq 10$\%~\citep{Sandell:94},  
was determined using K3-50.  
Atmospheric contributions to the signal were subtracted using 
a modified version of the algorithms described in 
\citet[][details in Wu et al. 2012]{Sayers:11}. 
The atmospheric subtraction also attenuates astronomical signal, and
to account for this, K3-50 was processed
in an identical way prior to determining the flux calibration.
   
The 1.1 mm~flux density at the location of \w1814 
is formally $-0.8$ mJy.
The noise fluctuations in the map are
Gaussian within our ability to measure them.
Including the estimated confusion noise rms per beam of 0.6~mJy,  
the estimated total rms per beam at the location of \w1814\ is 1.6~mJy.
Based on the statistical methods of \citet{Feldman:98}, 
the corresponding 95\% confidence upper limit is 2.4 mJy at 1.1~mm,  
as listed in Table~1. 

\subsubsection{Radio\label{sec:radio}}

\w1814\ was observed with 22 antennas of the D configuration of the
Expanded Very Large Array (EVLA) on 2010 June 10.  Observations were made
simultaneously in two 128 MHz bands centered on 
4.494~GHz (6.77~cm, ``C-band low") and 7.93~GHz (3.78~cm, ``C-band high") 
in order to obtain a spectral index.  The on-source
integration time was 29 minutes.  The source J1759+2343 was used for
phase calibration and 3C48 was used for flux calibration.
The C-band data were calibrated and imaged using 
Common Astronomy Software Applications\footnote{\tt http://casa.nrao.edu/} (CASA). 
The 4.494~GHz image has $39\ \mu$Jy/beam rms
noise in the $15.1\arcsec \times 12.3\arcsec$ beam (with major-axis PA
$= -42\deg$), and the 7.93~GHz image has $67\ \mu$Jy/beam noise 
with a $8.7\arcsec \times 7.1\arcsec$ beam at PA $= -37\deg$.

\begin{figure}
\plotfiddle{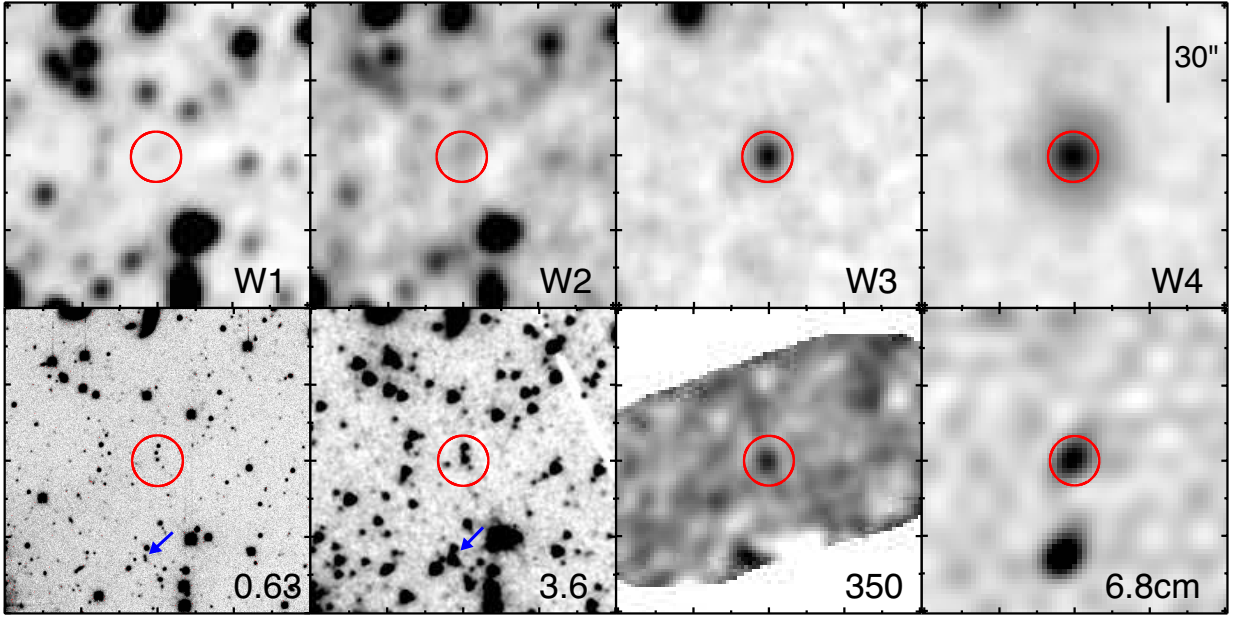}{5.0in}{0}{100}{100}{-235}{30}
\caption{Top row:
Images of \w1814\ in the WISE W1 (3.4 $\mu$m), W2 (4.6 $\mu$m), 
W3 (12 $\mu$m), and W4 (22 $\mu$m) bands, from left to right.  
Second row: Followup images at {\it r'} ($0.63\ \mu$m), 
$3.6\ \mu$m, $350\ \mu$m, and 4.5 GHz (6.8 cm), from left to right. 
Images are 2 arcmin on a side, with North up and East to the left.   
Circles are 10\arcsec\ in radius, centered on the WISE source position.
The counterpart to the southern radio source apparent in the 6.8 cm image 
is indicated with a blue arrow in the 0.63 and $3.6\ \mu$m images.
\label{fig:w12d}}
\end{figure}

Source positions and flux densities were estimated from Gaussian fits
made by the AIPS task IMFIT. An unresolved source coincident with
\w1814\ to within the astrometric errors 
(i.e. $\pm 0.03$ s in RA and $\pm 0\farcs5$ in dec)
and a flux density of $560 \pm 70\ \mu$Jy was detected 
at 4.494~GHz (Figure~\ref{fig:w12d}).  At 7.93~GHz, 
\w1814\ has a flux density of $350 \pm
130\mu$Jy, giving a spectral index $\alpha\approx 0.8$, where $F_\nu
\propto \nu^{-\alpha}$.  A second unresolved source was detected 4\arcsec\ east
and 39\arcsec\ south of \w1814\ with flux densities of $790 \pm70\ \mu$Jy at
4.494~GHz and $460 \pm120\ \mu$Jy at 7.93~GHz ($\alpha \sim 1.0$).  
The counterpart to this second
source at {\it r'} and [3.6] is indicated in Figure~\ref{fig:w12d}
with an arrow.  Using the spectral indices to
extrapolate the EVLA flux densities for the two sources to 1.4 GHz predicts
1.4 mJy for \w1814 and 2.4 mJy for the southern source. 
The sum of these flux densities is consistent with the NRAO VLA Sky Survey 
\citep[NVSS;][]{Condon:98} measurement of $3.4 \pm 0.5$~mJy, which is a blend of the two sources.

Observations of \w1814\ in the Ka band ($26-40$ GHz, or 9.7~mm at 32~GHz)
were acquired with the Green Bank Telescope (GBT) on 2010 November 10 using 
the Caltech Continuum Backend (CCB). 
A total on-source integration time of 25 minutes was obtained 
from 38 useful one-minute ``nod" (beam-switched) observations.  
The telescope pointing was corrected dynamically 
with reference to the bright nearby source 1814+4113.  
The flux calibration was established by an observation of 3C48, 
which was assumed to have a flux density at 32.0 GHz of 0.82 Jy, 
and a spectral index of $\alpha = 1.18$. 
This flux density is based on the 32 GHz brightness temperature measurement of Jupiter 
\citep[$T_J = 146.6 \pm 0.75 \, {\rm K}$;][]{Hill:09}.  
The data were reduced in publicly available IDL routines 
custom written to process GBT+CCB data, 
using the approach described in \citet{Mason:09}.  
All four 3.5 GHz channels over the 26 to 40 GHz frequency band 
were averaged together to maximize the sensitivity, 
giving a $2\sigma$ upper limit of $175\ \mu$Jy for \w1814.

\subsection{Optical Spectroscopy\label{sec:spectrum}}

We obtained the discovery spectrum revealing that \w1814\ was at $z=2.452$ 
on UT 2010 May 12 with the dual-beam Low Resolution
Imaging Spectrometer \citep[LRIS;][]{Oke:95} with its Atmospheric Dispersion Compensator (ADC) 
on the Keck~I telescope.  
Two dithered 600~s exposures were obtained in clear conditions 
with sub-arcsecond seeing. 
The observations used the 1\farcs5 wide longslit, the
6800 \AA\ dichroic, the 400 $\ell\, {\rm mm}^{-1}$ grating on the
red arm of the spectrograph (blazed at 8500 \AA; resolving power 
$R \equiv \lambda / \Delta \lambda \sim 700$ for objects filling
the slit), and the 400 $\ell\, {\rm mm}^{-1}$ grism on the blue arm
of the spectrograph (blazed at 3400 \AA;  $R \sim 600$).  
An offset star $22\farcs1\ $ to the northwest was used to position
the spectrograph slit on the WISE source, with
a slit position angle (PA) of 143.6\deg\ so that the offset star 
remained in the slit. 

Additional LRIS-ADC spectroscopy of \w1814\ and sources in
its vicinity (\S~\ref{sec:optical}, Figure~\ref{fig:nirc2}) was obtained
during UT 2010 July 13$-$15.  These observations used
the 1\farcs5 wide longslit, the 5600 \AA\ dichroic, the 600 $\ell\,
{\rm mm}^{-1}$ blue grism (blazed at 4000 \AA; $R \sim 1000$), 
and the same 400 $\ell\, {\rm mm}^{-1}$ red grating
as used in May 2010.  The nights were photometric.  
On UT 2010 July 14 two dithered 600~s exposures were obtained at a PA of 8.0\deg\  
to simultaneously observe components A and B, 
and two dithered 600~s exposures were obtained 
at a PA of 12.0\deg\ to simultaneously observe components C and D.  
On UT 2010 July 15 two dithered 900~s exposures were obtained 
at a PA of 205.0\deg\ to simultaneously observe components B and D.  
In each case, the red arm exposure time was 60~s shorter than the 
600~s or 900~s blue arm exposure time.  
The data were processed using standard procedures, and flux calibrated
using observations of standard stars from \citet{Massey:90}.
The reduced spectra are presented in Figure~\ref{fig:spectrum}.

The upper left panel of Figure~\ref{fig:spectrum} shows the July 2010 spectrum of 
component A, which is of slightly higher quality than the May 2010 data.  
\w1814\ shows a typical Lyman-break galaxy (LBG) spectrum 
at redshift $z = 2.452$, with Ly$\alpha$ emission, a continuum break across the
emission line due to the Ly$\alpha$ forest, and several UV 
interstellar absorption lines.  The spectrum shows no signs
of high ionization, high equivalent width emission lines typical
of AGN activity.  For comparison, the LBG composite spectrum from
\citet{Shapley:03} is shown directly below the component A
spectrum in Figure~\ref{fig:spectrum}.   
The May 2010 spectrum shows Ly$\alpha$ emission extending  
$>30$ kpc to the southeast (PA 143.6\deg) of component A, 
making it a Ly$\alpha$ blob \citep{Bridge:12}.

The upper right panel of Figure~\ref{fig:spectrum} shows 
the combined spectrum from the two PA's covering component B,   
which has a much redder optical
spectrum than the other components, and appears to be an early-type 
M-dwarf star.  The colors ${\it g'} - {\it r'} = 1.48$, ${\it J} - [4.5] = 0.95$, 
${\it K_s} - [3.6] = 0.11$,
and $[3.6] - [4.5] = 0.17$ support this interpretation
\citep{Bochanski:07,Patten:06}.  

The lower left panel of Figure~\ref{fig:spectrum} shows the spectrum of component~C. 
This source is a typical broad-lined (i.e., unobscured or Type 1) 
quasar at the same redshift
as \w1814.  Self-absorption is clearly evident in the Ly$\alpha$,
\ion{Si}{4} and \ion{C}{4} transitions.

The lower right panel of Figure~\ref{fig:spectrum} shows the spectrum of component~D. 
The spectrum has a relatively flat (in $F_\nu$) continuum,  
with a broad absorption and continuum break
at approximately the same wavelength as Ly$\alpha$ at $z \sim 2.45$.
Several interstellar absorption lines are evident that match this
interpretation, implying that component~D is at the same redshift
as \w1814.

\subsection{Astrometry\label{sec:astrometry}}

Table~2 gives the measured positions for \w1814\ components A, B, C, and D,
when the different components are detected.  Only the seconds
of right ascension and arcseconds of declination of the coordinates for components
B, C, and D are listed.  
For the WISE, CSO, and radio data, only a single component is detected.  
A minimum uncertainty of 0\farcs2 is assumed to account for systematic uncertainties 
with respect to an absolute reference frame, which is larger than the 
internal measurement errors for the optical, near-IR, 
and {\it Spitzer} data.   

\begin{figure}
\plotfiddle{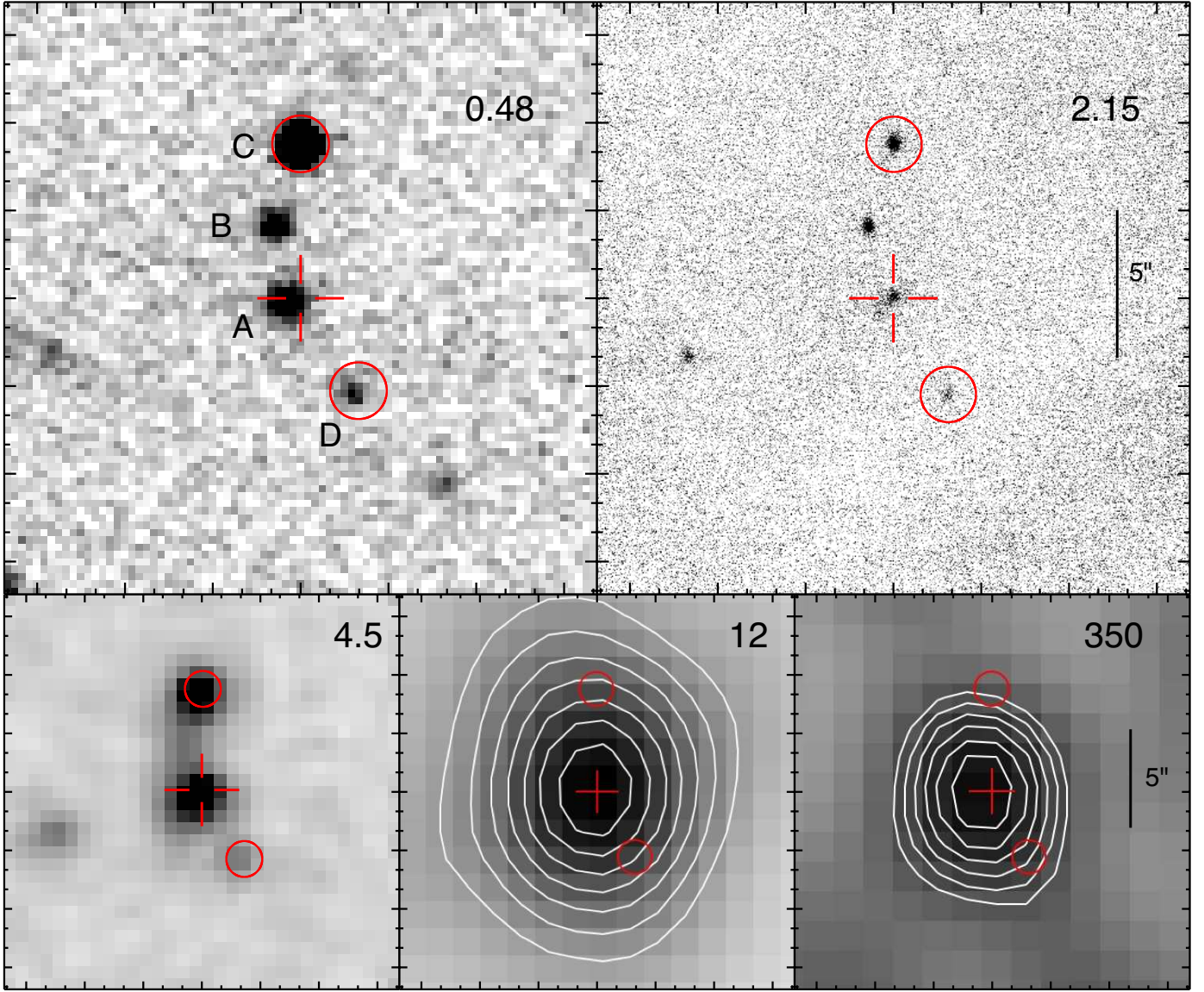}{5.7in}{0}{100}{100}{-265}{-5}
\caption{Images of the region near \w1814\ in {\it g'} ($0.48\ \mu$m - top left),
$K_s$ ($2.15\ \mu$m - top right), $4.5\ \mu$m - bottom left, 
W3 ($12\ \mu$m - bottom center), and $350\ \mu$m (bottom right).    
The field is 20\arcsec\ on a side with North up and East left.  
Component A is resolved in the {\it g'} and $K_s$ images, while B and C are unresolved.
Components A and D show a Lyman-break galaxy spectrum at $z=2.45$, component C
is a broad-lined quasar at $z=2.45$, and component B is an M-dwarf star.
The relative position in the $K_s$ image of component A with 
respect to component C is marked with red cross-hairs,
showing that component A is displaced $\sim 0\farcs5$ East in the {\it g'} image. 
Contours at 12 and $350\ \mu$m are from 30 to 90\% of the peak in each image. 
The emission at 12 and $350\ \mu$m is dominated by and centered on component A to within the 
astrometric errors, implying that the other $z=2.45$ components 
(C and D, whose locations are marked with circles)
make little contribution to the bolometric luminosity.
\label{fig:nirc2}}
\end{figure}

\begin{figure}
\plotfiddle{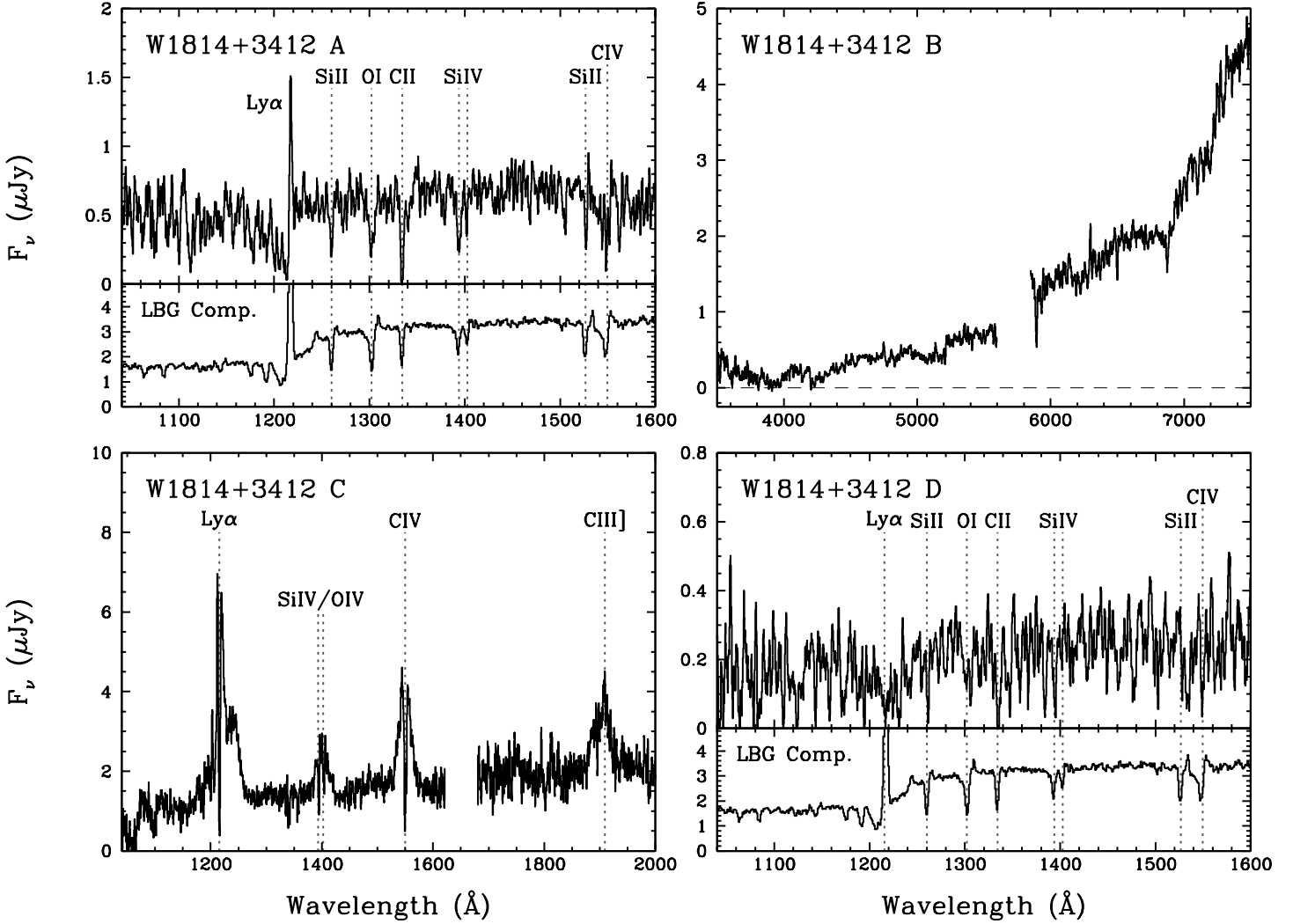}{0.0in}{270}{75}{75}{-300}{90}
\vspace*{12cm}
\caption{Keck/LRIS spectra of optical components near \w1814, as marked in
the upper left panel of Figure~\ref{fig:nirc2}. Upper left: Component~A,
shifted to the rest frame for $z=2.452$.
The composite rest-frame ultraviolet spectrum of $z \sim 3$ LBGs
from \citet{Shapley:03} is shown directly below for comparison.
Upper right: Component~B spectrum in the observed frame. 
This is an optically red point source with spectrum
and photometry consistent with an M-dwarf star. 
Lower left: Component~C in the rest frame for $z=2.452$, 
showing a bright quasar with
narrow self-absorption at the same redshift as \w1814.
Lower right: Component~D in the rest frame for $z=2.452$.  
The broad Ly$\alpha$ absorption, continuum break,
and detection of several interstellar absorption
lines imply that component~D is at the same redshift as \w1814.
The \citet{Shapley:03} composite LBG spectrum is again shown 
directly below for comparison.
\label{fig:spectrum}}
\end{figure}

As can be seen in the lower panels of Figure~\ref{fig:nirc2}, 
the positions of component A and of the single detected component in the longer wavelength data
are in agreement, within the errors.\footnote{The WISE catalog coordinates
listed in Table~2 are in agreement with the coordinates for component A in other bands,
and have a systematic uncertainty with respect to 2MASS of $<0\farcs2$.
The world coordinate system (WCS) in WISE image tiles deviates slightly 
but systematically from the catalog coordinates for sources in the tile, 
with the catalog coordinates being correct.   
For the image tile containing \w1814\ this offset
causes sources to appear 0\farcs68 East and 1\farcs10 South 
of their overlaid catalog coordinates, and we have adjusted the WISE image WCS
to correct for this in the bottom center panel of Figure~\ref{fig:nirc2}.}  
With a reduced chi-squared value of 0.984 for the
profile fit to a point source in the WISE data, \w1814\ is unresolved. 
In W3, with a reduced chi-squared of 0.98, the source FWHM is $< 1$",  
while in W4 the reduced chi-squared of 1.09 corresponds to a FWHM $< 6$".  
The separations of 5\farcs2 between component A and C, 
and of 3\farcs8 between component A and D, are large compared
to the uncertainties in the WISE position, 
or in the 4.5 GHz (C-band low) position.  
We therefore conclude that component A at $z=2.452$
is predominantly responsible for the flux from the single component seen in the longer wavelength data, and hereafter refer to component A as \w1814\ unless otherwise indicated.

However, Figure~\ref{fig:nirc2} suggests that there is a noticeable shift 
in the position of component A with respect to component C in the highest resolution
images.  Component A lies essentially due south of
component C in the $K_s$ and \spitzer\ images, but 
in the optical images component A is shifted $\approx 0\farcs5$ 
to the east, or 4 kpc at $z=2.452$.  
As a point source and $z=2.45$ quasar, component C 
is almost certainly at a fixed position in all images, so we
believe this shift in the component A optical centroid position 
with respect to the $K_s$ and \spitzer\ positions to be real.
Because the optical spectrum and photometry 
suggest appreciable extinction is present (\S\ref{sec:SFR}),  
we consider it most likely that the $K_s$ and \spitzer\ 
positions of component A are coincident with the WISE and radio positions,
with the optical centroid of component A slightly displaced from these.

\section{Analysis}

\begin{figure}
\plotone{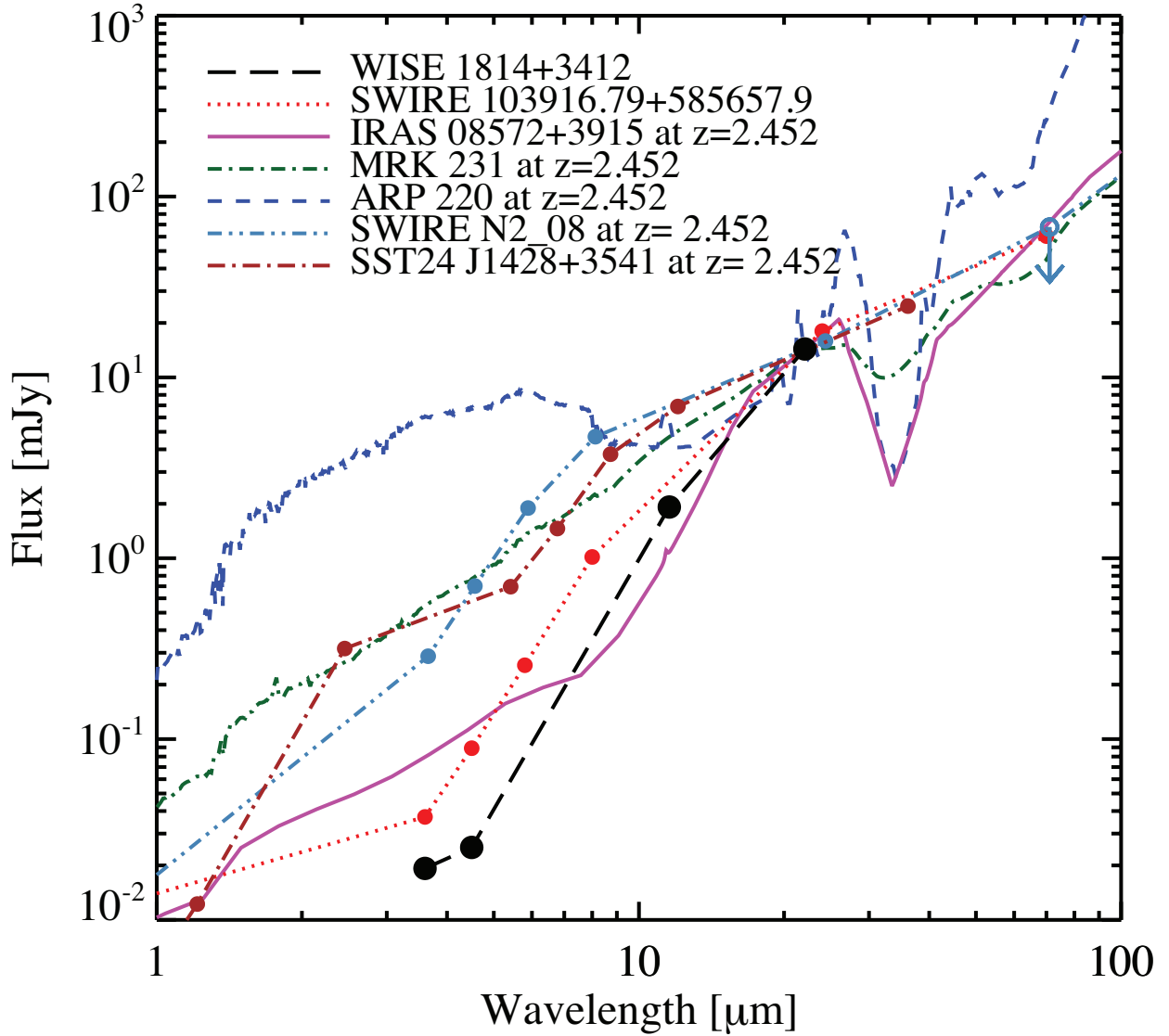}
\caption{\spitzer\ photometry at 3.6 and $4.5\ \mu$m
and WISE photometry at 12 and $22\ \mu$m for
\w1814, compared to SED's of other unusually red objects
whose wavelengths have been scaled to $z=2.452$ and flux
densities scaled to the $22\ \mu$m flux density of \w1814. 
The approximate flux density scaling factors and observed redshifts are:
SWIRE J103916.79+585657.9 \citep[4, $z$ unknown but assumed to be 2.452,][]{Weedman:06};
IRAS 08572+3915 (0.06, $z=0.058$); 
Mrk 231 (0.02, $z=0.042$); 
Arp 220 (0.05, $z=0.018$);  
SWIRE N2\_08  
\citep[SWIRE J164216.93+410127.8, 4, $z=2.40$,][]{Polletta:08}; 
and SST24 J1428+3541 \citep[2.3, $z=1.293$,][]{Desai:06}.  
\label{fig:sedredobjects}}
\end{figure}

Figure~\ref{fig:sedredobjects} compares WISE and {\it Spitzer} photometry for
\w1814\ to other unusually red objects redshifted to $z=2.452$ and normalized at 22 $\mu$m.  
These include the well known ULIRGs Arp 220 and Mrk 231; 
the extremely red $z=0.058$ IRAS source 08572+3915;  
the $z=1.293$ source SST24 J1428+3541 \citep{Desai:06},
which is the brightest $24\ \mu$m source (10.55 mJy) among mid-IR 
sources with faint optical counterparts in the $8\ {\rm deg}^2$ Bo\"otes field;  
the $z=2.40$ source N2\_08 (SWIRE J164216.93+410127.8),
which is the most luminous obscured QSO from the sample of \citet{Polletta:08};  
and source A10 (SWIRE J103916.79+585657.9) from \citet{Weedman:06},
which does not have a measured redshift and is among the reddest sources found in the SWIRE survey.
None of these are as extreme as \w1814 in the W2 to W3 range (4.6 to $12\ \mu$m).
\citet{Wu:12} show that W1W2-dropouts readily satisfy the $R - [24] \ge 14$ 
criterion for ``DOGs" \citep[dust obscured galaxies,][]{Dey:08}.
\citet{Dey:08} suggest the $24\ \mu$m emission in the brighter DOGs arises from
warm dust heated by an AGN, and since the $24\ \mu$m flux density limit for DOGs 
is $\sim 20\times$ fainter than for W1W2-dropouts, AGN might be expected to play an 
important role in W1W2-dropouts. 

\subsection{SED Modeling\label{sec:SED}}

\begin{figure}
\plotone{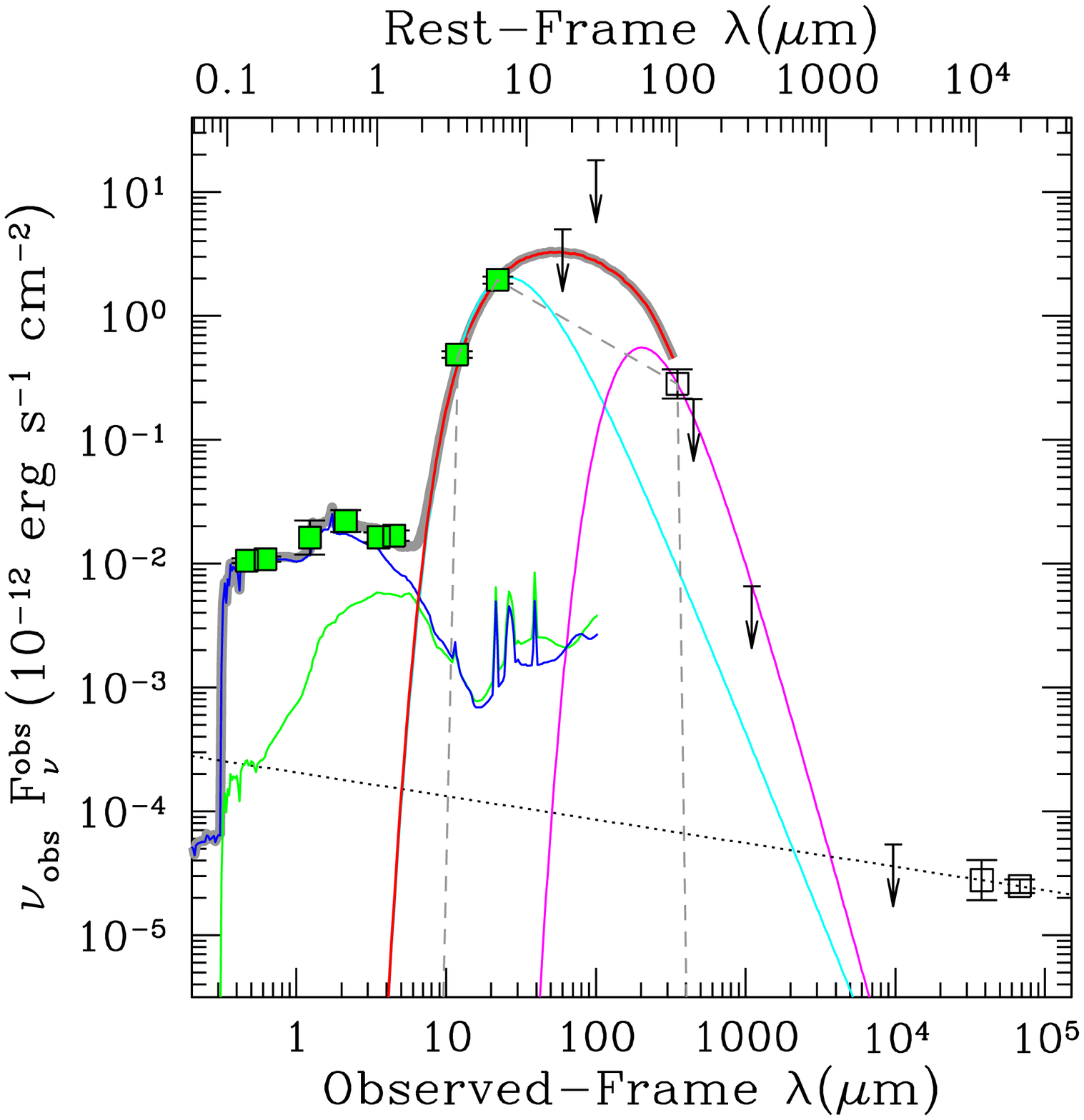}
\caption{Best fit SED template model (heavy grey line) 
to the photometry (shown by green symbols)
for \w1814\ plotted in $\nu F_\nu$ units. 
Other measurements (black symbols) and upper limits (black downward pointing arrows)
were not used for the template fit.
The template fit uses starburst (blue) and Sbc spiral (green) components from \citet{Assef:10}, 
and the \citet{Richards:06b} Type I AGN template (red) with $A_V = 48$ extinction applied.  
The cyan line shows a 
488 K blackbody fit to the W3 ($12\ \mu$m) and W4 ($22\ \mu$m) photometry which overlaps the
red curve below $12\ \mu$m,
while the magenta line shows a modified 
(i.e., with dust emissivity $\propto \nu^{1.5}$) 45 K blackbody fit to 
the $350 \mu$m flux density and the upper limit at 1.1 mm. 
The dotted line shows an $F_\nu \propto \nu^{-0.8}$ power law fit to the radio data.
The grey dashed line shows the power laws used to estimate the minimum 
bolometric luminosity in \S\ref{sec:luminosity}.
\label{fig:sedmodel}}
\end{figure}

Figure~\ref{fig:sedmodel} plots the observed photometry for
\w1814.  
The detections in W3 (12 $\mu$m), W4 (22 $\mu$m), and $350\ \mu$m and the upper limits
at $450\ \mu$m and 1.1 mm imply the luminosity (\S~\ref{sec:luminosity}) 
is dominated by emission from warm dust, possibly heated by an AGN.  
The EVLA detections show an AGN is present (\S~\ref{sec:radioFIR}).  
The similarity of the \w1814\ optical spectrum to 
a Lyman Break Galaxy (\S~\ref{sec:spectrum}) implies that
active star formation is underway, as discussed in
\S~\ref{sec:SFR}. 

Motivated by these components, in Figure~\ref{fig:sedmodel} 
we show an example fit to the optical, near-IR, \spitzer\, and WISE photometry for \w1814\,  
using starburst and evolved stellar population templates from \citet{Assef:10}, 
and a reddened Type 1 AGN template from 
\citet{Richards:06b}.\footnote{
Due to its longer wavelength range, the \citet{Richards:06b}
AGN template was used rather than that from \citet{Assef:10}, 
although the fit is qualitatively identical with either template.}  
As in \citet{Assef:10}, the reddening is assumed to follow
an SMC-like extinction curve \citep{Gordon:98} for $\lambda < 3300$ \AA\ 
and a Galactic extinction curve \citep{Cardelli:89} at longer wavelengths, 
in both cases with $R_V = 3.1$.   The resultant AGN template extinctions 
in the rest frame are $E(B - V) = 15.6 \pm 1.4$ or $A_V = 48 \pm 4$, 
corresponding to $A_{W3} = 2.7 \pm 0.2$ and $A_{W4}=0.98 \pm 0.09$
in the observed frame.  
Because the \citet{Richards:06b} template does not extend to $350\ \mu$m ($100\ \mu$m rest),
this reddened Type 1 AGN component was {\em not} fit to the submillimeter photometry, 
but with a small extrapolation it nevertheless provides a good match to those data. 
The cyan line in Figure~\ref{fig:sedmodel} shows a 
490 K blackbody fit to the W3 ($12\ \mu$m) and W4 ($22\ \mu$m) photometry,
while the magenta line shows a modified 
(i.e., with dust emissivity $\propto \nu^{1.5}$) 45 K blackbody fit to 
the $350 \mu$m flux density and the upper limit at 1.1 mm. 
Finally, the dotted line shows a radio spectral index $\alpha = 0.8$
fit to the EVLA data (\S~\ref{sec:radio}).  We consider each of these
components in more detail below.

\subsection{Luminosity Estimates\label{sec:luminosity}}

We obtain a lower limit of $L_{\rm bol} > 3.7 \times 10^{13} L_\odot$ 
for the bolometric luminosity of \w1814\ 
by assuming the SED is given by two power laws
connecting the observed flux densities at W3, W4, and $350\ \mu$m, with
no luminosity at shorter or longer wavelengths, as shown by the light dashed
lines in Figure~\ref{fig:sedmodel}. 
The luminosity can also be estimated 
by integrating over the best-fit SED templates of \S~\ref{sec:SED},
which are dominated by the dust-reddened AGN component, extending the latter  
beyond $350\ \mu$m via a $F_\nu \propto \nu^{3.5}$ power law. 
The details of this extrapolation are not significant as 
nearly all of the flux is already contained in the reddened AGN template. 
This approach yields a bolometric luminosity estimate of
$9.2 \pm 1.0 \times 10^{13} L_{\odot}$, of which 
$7.2 \pm 0.8 \times 10^{11} L_{\odot}$ ($<1\%$) is from the stellar components.   
From the upper limit to the IRAS 60 $\mu$m flux density (\S\ref{sec:IRAS})
plotted in Figure~\ref{fig:sedmodel}, the luminosity is unlikely to be a factor of two higher.  

\subsubsection{Lensing\label{sec:lensing}} 

Luminosities approaching $10^{14} L_{\odot}$
inevitably lead to suspicions that 
\w1814\ is gravitationally lensed,
as is true for sources such as
IRAS~FSC10214+4724 \citep[e.g.,][]{Eisenhardt:96} 
and APM 08279+5255 \citep[e.g.,][]{Ibata:99a}.
However, Figures~\ref{fig:nirc2} and \ref{fig:spectrum}
show lensing is unlikely to be a significant factor
for \w1814.  Although there are multiple components
at the same redshift (A, C, D), their SEDs and spectra 
have different characteristics.
The obvious candidate for a foreground lens is
component~B, but this appears to be a Galactic M dwarf star (\S~\ref{sec:spectrum})
rather than a massive galaxy at intermediate redshift.  
The extended $K_s$ emission at PA$\sim140\deg$ around component A seen in Figure~\ref{fig:nirc2} 
is unlikely to come from a foreground lensing galaxy, as it coincides 
with extended $z=2.452$ Ly$\alpha$ emission at the same PA (\S~\ref{sec:spectrum}).  
Therefore the large luminosity of \w1814\ appears to be intrinsic.

\subsection{Starburst Component\label{sec:SFR}}

The spectrum of \w1814\ is typical of an LBG (\S~\ref{sec:spectrum}), 
implying active star formation directly detected in the rest-frame UV. 
Assuming no UV extinction, the star formation rate (SFR) from the 
average $L_\nu$ of the starburst component of our fit (\S~\ref{sec:SED})
over rest 1500 - 2800 \AA\ and equation (1) of \citet{Kennicutt:98}
is $\sim 50 M_\odot {\rm yr}^{-1}$ for a Salpeter initial mass function (IMF) from 0.1 to $100\ M_\odot$, 
or $\sim 30 M_\odot {\rm yr}^{-1}$ for a \citet{Chabrier:03} IMF. 

However, even UV-selected LBG galaxies show evidence for extinction \citep{Shapley:03}. 
We estimate the extinction in \w1814\ by comparing the observed $g' - r' = 0.34$ color (AB system)
to the color expected from \citet{Bruzual:03} models, 
using the ``Ez\_Gal" web tool\footnote{\tt http://www.mancone.net/ezgal/model}.
Continuously star-forming models with Salpeter and Chabrier IMF's, 
solar and 0.4 solar metallicities, and ages of 0.1 to 0.4 Gyr were considered.  
The model $g' - r'$ colors range from $0.01$ for solar metallicity models which have
been forming stars for 0.4 Gyr, to $-0.12$ for 0.1 Gyr models with 0.4 solar metallicity,
and do not change substantially between the two IMF's. 
Using the reddening law from equation (4) of \citet{Calzetti:00},
these colors imply extinction at $0.16\ \mu$m ranging from 2.00 to 2.75 mag.
The corresponding corrected SFR values are $\sim 300 (180)$ to $600 (350)\ M_\odot {\rm yr}^{-1}$ 
for a Salpeter (Chabrier) IMF.  Higher SFR values are possible if some is completely obscured in the UV.

\subsection{Stellar Component\label{sec:stars}}

The stellar mass can be estimated from the observed 17.13 magnitude at 4.5 $\mu$m, 
the longest wavelength in the SED dominated by the stellar component
(Figure~\ref{fig:sedmodel}). The starburst component of the template fit 
(\S~\ref{sec:SED}) falls well below the observed $K_s$, [3.6] and [4.5] SED,
requiring a significant contribution from older stars, as represented by the Sbc template.

An upper limit for the plausible stellar mass 
can be obtained using the \citet{Bruzual:03} models 
by assuming that the observed [4.5] flux density is from stars formed 
in a 0.1 Gyr burst at $z=10$ (i.e. 2 Gyr before they are observed at $z=2.452$)  
with a Salpeter IMF and solar metallicity.  This model predicts $M/L=0.51$ in rest {\it K},
much higher than the $\sim 0.1$ value typical of the starburst models considered in \S~\ref{sec:SFR},
but still well below the value of $\sim 1$ typical of present day galaxies \citep{Bell:03}.  
Allowing for extinction of 0.34 mag at [4.5] corresponding to the upper end
of the range inferred in \S~\ref{sec:SFR} from the observed vs. model $g' - r'$ colors, 
this approach gives an upper limit to the stellar mass of $5.8 \times 10^{11} M_\odot$.
While larger extinction values and hence masses are of course possible, 
this value is already many times the stellar mass of a field L* galaxy today
\citep[$\sim 5 \times 10^{10} M_\odot$; e.g. ][]{Baldry:08}.

A lower limit can be obtained assuming that all the observed [4.5] light comes
from one of the starburst models considered in \S~\ref{sec:SFR}.  The smallest stellar mass
obtained with this approach is $4.1 \times 10^{10} M_\odot$, 
for a 0.1 Gyr old burst with solar metallicity and a Chabrier IMF.
A more plausible model with equal contributions to the [4.5] luminosity from a starburst
and an evolved stellar population, and with a more typical extinction at [4.5] of 0.3 mag, 
yields a stellar mass of $3.6(2.0) \times 10^{11} M_\odot$, for a Salpeter/Chabrier IMF.

\subsection{Dust Component\label{sec:dust}}

The prominent emission of \w1814\ in W3 and W4 implies 
a significant contribution from warm dust.  
Fitting a blackbody 
to the W3 to W4 flux density ratio of 0.13 at $z=2.452$ yields a temperature
of $\sim490$ K (cyan curve in Figure~\ref{fig:sedmodel}).  
Of course a range of temperatures is needed to account for all the data, 
but a modified blackbody (with $F_\nu \propto \nu^{1.5} B_\nu$)
fit to the W4 to $350 \mu$m ratio of 0.44 
also implies warm dust ($\sim 170$ K).
The scale of the warm dust is substantial: 30 to 250 pc for the radius of 
a spherical blackbody with $L_{\rm bol} = 9 \times 10^{13} L_\odot$  
and temperatures of 490 to 170 K. 
Using equation (1) of \citet{Calzetti:00} with $F_{350}/(1+z)$ as the rest $100\ \mu$m flux density, 
the warm dust mass is $\simgt 10^7 M_\odot$. 
The gas mass is $\simgt 10^9 M_\odot$ for the Milky Way gas to dust ratio, 
but could be up to $\sim 10^{11}M_\odot$ using the gas to dust ratio in AGN \citep{Maiolino:01}.

In contrast, the ratio of the $350\ \mu$m to 1.1 mm flux density is greater than 13.75, 
which means that dust at the cooler temperatures associated with sub-mm galaxies
\citep[$\sim 35$ K; e.g.][]{Kovacs:06b}
plays a smaller role in the bolometric luminosity of this source.
\citet{Wu:12} find that this predominance of warm dust is a characteristic of W1W2-dropouts, 
using CSO observations of 26 sources. 
For \w1814, dust cooler than 45 K with $F_\nu \propto \nu^{1.5} B_\nu$ 
normalized to the observed $350\ \mu$m flux density
would exceed the 2.4 mJy 95\% confidence upper limit at 1.1 mm
(magenta curve in Figure~\ref{fig:sedmodel}).  

Nevertheless it seems likely that \w1814\ contains a cooler dust component.
Figure~\ref{fig:nirc2} (see also \S\ref{sec:astrometry})
shows the rest UV emission from component A is displaced from the 
rest optical and near-IR by $\sim 4$ kpc, suggesting extinction
on a much larger scale than the warm dust.  
The $L_{IR}$ expected from a $\sim 300\ M_\odot {\rm yr}^{-1}$ SFR (\S\ref{sec:SFR})
is $\sim 2\times10^{12}L_\odot$ \citep{Kennicutt:98}. 
Dust with this luminosity at 35 K with $F_\nu \propto \nu^{1.5} B_\nu$ would 
contribute more than half the $350\ \mu$m flux density and just satisfy the 1.1 mm limit, 
and have a mass of $\sim 2.5 \times 10^{8} M_\odot$.

\subsection{Radio Source\label{sec:radioFIR}}

The rest 1.4 GHz luminosity is $\sim 5 \times 10^{25}\ {\rm W\ Hz^{-1}}$,
and the ratio of rest 5 GHz ($\sim$observed 1.4 GHz) to rest 4400 \AA\ ($\sim$ observed $J$)
flux densities is $\sim 200$. Both metrics qualify \w1814\ as radio-loud \citep{Stern:00a}.   
If free-free emission from HII regions associated with star formation 
were present in \w1814\ at the level seen in M82
\citep[][Figure~1]{Condon:92}, the Ka band upper limit
of $175\ \mu$Jy limit (\S~\ref{sec:radio}) should have been exceeded,
but the limit is consistent with the $115\ \mu$Jy flux density expected 
from extrapolating the EVLA observations.  
This argues against star formation dominating the energetics of \w1814.  

Using the template SED fit and radio properties to
estimate the $z=2.452$ rest frame flux densities at $60\ \mu$m, $100\ \mu$m 
and 1.4 GHz (90, 33 mJy, and 3.9 mJy respectively), we obtain $q=1.36$ 
from equations 14 and 15 of \citet{Condon:92}. This is
$\sim10\times$ more radio loud than the normal value of $2.4\pm0.24$ \citep{Ivison:10},
again implying the presence of an AGN. 
We conclude the radio emission in \w1814\ is due to AGN activity, 
rather than to star formation.

\subsection{AGN Component\label{sec:AGN}}
 
Over 99\% of the bolometric luminosity in the template fit to the SED 
comes from the reddened Type 1 AGN component (\S\ref{sec:luminosity}),
although there are no AGN features in the component A spectrum (\S\ref{sec:spectrum}). 
\citet{Wu:12} show W1W2-dropouts exhibit a range of rest-UV spectra, 
from an absence of any AGN signatures to the presence of broad lines, 
with most showing AGN features.  The most direct evidence for an AGN in \w1814\ 
comes from the radio emission, which is above the radio-IR correlation (\S\ref{sec:radioFIR}). 
Since the SED modeling finds $A_V=48$ mag, corresponding to $\sim 200$ mag extinction 
in the rest UV, it is not surprising that AGN features such as broad lines and UV continuum 
from close to the black hole are absent from the spectrum.  
The corresponding proton column to the AGN is $\sim 10^{23}\ {\rm cm}^{-2}$ for Milky Way reddening 
\citep{Bohlin:78}, but up to $\sim 10^{25}\ {\rm cm}^{-2}$ (i.e. Compton-thick)
using gas to dust ratios for AGN \citep{Maiolino:01}.   
If the source hosts an extremely luminous and obscured AGN, data from approved
XMM-Newton and NuSTAR \citep{Harrison:10} programs should reveal  
a hard X-ray spectrum, with few photons below rest-frame 10 keV.
  
Deep spectropolarimetry might detect AGN features.  Such work has been done for example
for $z\sim2$ radio galaxies with similar optical brightness,  
with exposures typically needing most of a Keck night per target \citep{Dey:96, Cimatti:98},  
and IRAS~FSC10214+4724 \citep[][optically much brighter than \w1814]{Goodrich:96}. 
All these objects show high ionization, narrow emission lines in their unpolarized spectra, and 
the lack of such features in \w1814\ is interesting, since typically they come from regions 
on a kpc scale \citep[e.g.,][]{Bennert:02}.
This suggests that either the narrow line region is more compact for this source 
and/or that obscuring material is present on larger scales than 
the torus \citep[e.g.,][]{Brand:07}.  The shift between the rest-UV and optical continuum
(Figure~\ref{fig:nirc2}) provides support for this idea.

If the luminosity in \w1814\ is due primarily to accretion 
onto a super massive black hole (SMBH), the SMBH mass for the template fit AGN luminosity 
is $2.8 \times 10^9 (L_{\rm Edd}/L_{\rm bol}) M_\odot$, 
where  $(L_{\rm bol}/L_{\rm Edd})$ is the Eddington ratio.   
The dust sublimation radius is $\sim 7$ pc,  
and the outer radius of the dust torus is expected to be no more than $30 \times$ 
the sublimation radius \citep{Nenkova:08}.  
This size is consistent with the range discussed in \S\ref{sec:dust}, but larger scale dust 
would be needed to obscure a kpc scale narrow line region.

\section{Summary and Discussion}

With a minimum bolometric luminosity of $3.7 \times 10^{13} L_\odot$,
and more likely $L_{\rm bol} \sim 9 \times 10^{13} L_\odot$,  
\w1814\ easily qualifies as a 
hyper-luminous infrared galaxy,
the first identified by WISE.  Its W4 (22 $\mu$m) flux density and 
redshift exceed that of the brightest $24\ \mu$m source DOG in the
$8\ {\rm deg}^2$ Bo\"otes field \citep[SST24 J1428+3541;][]{Desai:06}, 
giving it a rest $5\ \mu$m luminosity $3\times$ higher.  
Its luminosity probably
exceeds the estimated $4\times 10^{13} L_\odot$ of the most luminous
obscured quasar in \citet[SWIRE J164216.93+410127.8;][]{Polletta:08},
which has $4\times$ lower flux at $24\ \mu$m. 
It approaches the $\sim 2 \times 10^{14} L_\odot$ of the most luminous known quasars 
such as S5~0014+813 \citep{Kuhr:83}, HS~1700+6416 \citep{Reimers:89}, and
SDSS~J074521.78+473436.2 \citep{Schneider:05}.  Several W1W2-dropouts
reported in \citet{Wu:12} are even more luminous than \w1814, 
with {\em minimum} $L_{\rm bol}$ up to $1.8\times10^{14} L_\odot$.

\w1814\ is also a massive, actively
star forming LBG, with an extinction corrected UV star formation rate of
$\sim 300\ M_\odot {\rm yr}^{-1}$ and a stellar mass of
$\sim 3 \times 10^{11} M_\odot$ 
\citep[several times the stellar mass of a field
$L*$ galaxy today;][]{Baldry:08}.  The specific
star formation rate of $\sim 1\ {\rm Gyr}^{-1}$ places
\w1814\ in the starburst galaxy regime, but even the upper end of
the extinction corrected UV SFR estimates account
for $\simlt 10\%$ of the bolometric luminosity. 
With no AGN signatures in the UV spectrum, 
it is tempting to assume that additional dust-obscured star formation
accounts for the high bolometric luminosity.  However, the upper limit at 1.1 mm 
would be greatly exceeded unless the star forming dust was  
significantly warmer than 35 K.

This possibility merits consideration.
\citet{Kovacs:06b} present evidence that starburst dominated sub-millimeter galaxies  
have higher temperatures at higher redshifts and luminosities.
The SED of the extreme low metallicity dwarf starburst galaxy SBS 0335-052 
peaks at $30 \mu$m \citep{Houck:04}, and \citet{Hirashita:04} suggest that 
virtually all high redshift star formation may occur in the ``active mode" 
for which SBS 0335-052 is the prototype.   The active mode is
characterized by dense, compact star-forming regions in which dust
from core collapse SNe warms as it screens the gas from UV photons, 
enabling runaway star formation as the gas continues to cool on a dynamical timescale. 
Yet it is not obvious that this mechanism can be scaled up to the levels   
required to produce $\sim 10^{14} L_\odot$,
i.e. a SFR $\sim 10^4 M_\odot {\rm yr}^{-1}$ and $\sim 10^8$ O stars. 
  
Since \w1814\ contains a modestly powerful radio source, 
an active galactic nucleus is almost certainly present. 
An obscured AGN powered by a super massive black hole (SMBH)
could supply the luminosity while being consistent with 
the absence of AGN signatures in the UV spectrum.  
In fact, despite the clear starburst signature from the optical spectrum, 
the lack of submm emission and predominance of hot dust suggest that 
in \w1814\ \citep[and other W1W2-dropouts;][]{Wu:12} we may be seeing an SED 
more like the ``intrinsic" SED from the dusty central AGN region than 
is the case for many quasars \citep[see Figure 6 of][]{Netzer:07}.
 
The formation and coevolution of AGN and stellar populations 
is the subject of much current research. 
Theoretical scenarios suggest the initial starburst 
rapidly enshrouds a forming galaxy in cold dust which emits at submm wavelengths.  
This is followed by an increase in fueling of the SMBH, 
triggering an AGN, generating warmer dust and mid-IR emission.  
Finally outflows from the AGN or starburst clear the dust and gas, 
revealing an optical quasar, and then removing the fuel which powers 
both starburst and quasar, leaving a quiescent massive galaxy 
\citep[e.g.,][]{Sanders:88, Hopkins:06, Narayanan:09}.   

The black hole mass needed to power the bolometric luminosity is 
$2.8 \times 10^9 (L_{\rm Edd}/L_{\rm bol}) M_\odot$. 
But using the local $L_{\rm bulge}/M_{\rm BH}$ relation, 
the SED fit stellar component absolute magnitude of $M_K = -26.28$   
gives a black hole mass of only $1.4 \times 10^9 M_\odot$ \citep{Graham:07}.
If the relation is actually between SMBH mass and bulge {\em mass} 
rather than bulge luminosity, as seems plausible, the discrepancy is larger.  
At $z=2.452$ the stellar mass for a given $K$ luminosity 
is likely less than half as large as today. 
Therefore, assuming the bolometric luminosity of the AGN template fit is correct, 
and that the \citet{Graham:07} relation 
adjusted for lower $(M/L)_K$ at $z=2.45$ applies, 
the implied Eddington ratio is $\simgt 4$.
For comparison, \citet{Kollmeier:06} measure a typical
Eddington ratio in luminous quasars of 0.25, and \citet{Kelly:10} find 0.1
to be representative.  If \w1814\ has a similar Eddington ratio, then one
of these assumptions does not hold.  

The local stellar mass to SMBH mass relation could apply
if the bolometric luminosity is mainly from star formation with warm dust similar to that in 
low metallicity starburst dwarf galaxies, but scaled up a factor of a thousand.
There is evidence that starbursts can 
dominate luminosities even in sources with spectral signatures of quasars 
\citep[e.g.,][]{Netzer:07, Polletta:08}. 
Alternatively, the stellar mass may have been underestimated by a factor of several
due to greater extinction at $4.5\ \mu$m,  but the extinction corrected
stellar mass of $>10^{12} M_\odot$ is then extreme. 
Note this argument holds for AGN with $L > 10^{14} L_\odot$ in general, and
the largest SMBH masses in the most luminous objects may be well above
the normal relation \citep[e.g.,][]{McConnell:12, Salviander:12}. 
Another possibility is that \w1814\ is in an unstable transitional stage 
where the Eddington ratio is well above 1.  Recently, \citet{Kawakatu:11} have argued
that SMBH's with super-Eddington accretion will be characterized by 
a low ratio of near-IR to bolometric luminosity, which is essentially
the selection criterion for \w1814.

We consider it more likely that \w1814\ lies well off the local 
$M_{BH}$-bulge mass relation, suggesting it may be in an early stage of its evolution 
where AGN activity is dominant and the object has yet to build 
the bulk of its stellar mass.  This would, however, contradict 
the expectation that AGN activity should lag 
the peak of star formation \citep[see e.g.,][]{Hopkins:12}.
In this context, the presence of 3 systems (components A, C, and D) at the
same redshift and within 50 kpc suggests that a significant
amount of stellar mass assembly lies in \w1814's future.

\acknowledgements

The authors thank Alex Pope for insightful discussions regarding the dust content
of \w1814; Alice Shapley for assistance in determining the star formation
rate associated with a Lyman Break Galaxy; 
R.~S.~McMillan, J.~V.~Scotti, J.~A.~Larsen, and G.~J.~Bechetti for early ground-based
followup observations; 
Conor Mancone for providing and answering questions about the 
convenient ``Ez\_Gal" web interface to the Bruzual-Charlot models; 
Leonidas Moustakas for suggesting references on the mass of L* galaxies; 
Jeonghee Rho for allowing observations of \w1814\ during her CSO time; 
and Tom Soifer for allocating \spitzer\ Director's Discretionary Time to observe \w1814; 
and the anonymous referee for suggestions which improved the presentation of the paper.  

R.J.A. and J.W. were supported by an appointment to the NASA Postdoctoral Program 
at the Jet Propulsion Laboratory, administered by Oak Ridge Associated 
Universities through a contract with NASA.
NJE acknowledges support from NSF Grant AST-1109116. 
This publication makes use of data products from the Wide-field Infrared Survey Explorer, 
which is a joint project of the University of California, Los Angeles, 
and the Jet Propulsion Laboratory/California Institute of Technology, 
funded by the National Aeronautics and Space Administration.
This work is based in part on observations made with the {\it Spitzer Space
Telescope}, which is operated by the Jet Propulsion Laboratory,
California Institute of Technology under contract with NASA.  
Some of the data presented here were obtained at the W.M. Keck Observatory, 
which is operated as a scientific partnership among  
Caltech, the University of California and NASA. 
The Keck Observatory was made possible by the generous financial support 
of the W.M. Keck Foundation.  
Some of the data are based on observations obtained at the Hale Telescope, 
Palomar Observatory as part of a continuing collaboration between the 
California Institute of Technology, NASA/JPL, and Cornell University.
Some of the data presented here were obtained at  the
Kitt Peak National Observatory, National Optical Astronomy Observatory, 
which is operated by the Association of Universities for Research in Astronomy 
(AURA) under cooperative agreement with the National Science Foundation.  
The National Radio Astronomy Observatory is a facility of the 
National Science Foundation operated under cooperative agreement by Associated Universities, Inc. 

Facilities:  
\facility{Wide-field Infrared Survey Explorer, 
{\it Spitzer Space Telescope} (IRAC) (MIPS), 
Palomar 200''(WIRC) (LFC), 
Keck (LRIS), Keck (NIRC2), 
KPNO (Mosaic), 
CSO (SHARC II) (Bolocam), 
EVLA, GBT}


\bibliographystyle{apj}
\bibliography{apj-jour,thebibliography}

\clearpage

\begin{deluxetable}{ccccc}
\tabletypesize{\normalsize}
\tablecaption{Photometry for \w1814\ 
\label{table:phot}}
\tablewidth{0pt}
\tablehead{
\colhead{Band} & \colhead{$\lambda (\mu$m)} & \colhead{Instrument} & \colhead{Flux Density ($\mu$Jy)} &
\colhead{Comp. B ($\mu$Jy)}}
\startdata
{\it g'} & 0.48 & MOSAIC & $1.64\pm0.074$ & $0.81\pm0.030$ \\
{\it r'} & 0.63 & MOSAIC & $2.25\pm0.11$ & $3.16\pm0.063$ \\
{\it J} & 1.25 & NIRC2 & $6.6\pm2.1$ & $14.0\pm3.0$ \\
$K_s$ & 2.15 & NIRC2 & $15.7\pm3.2$ & $10.9\pm2.2$\\
$[3.6]$ & 3.6 & IRAC & $19.3\pm1.9$ & $5.06\pm0.49$\\
$[4.5]$ & 4.5 & IRAC & $25.2\pm2.4$ & $3.77\pm0.39$\\
W3 & 12 & WISE & $1860\pm100$ & \\ 
W4 & 22 & WISE & $14380\pm870$ & \\
& 60 & IRAS & $<100000$ & \\
& 100 & IRAS & $<600000$ & \\
& 350 & SHARC-II & $33000\pm9000$ & \\
& 450 & SHARC-II & $<35000$ & \\
& 1100 & Bolocam & $<2400$ & \\
Ka & 9700 & GBT & $<175$ & \\
C-band high & 37800 & EVLA & $350\pm130$ & \\
C-band low & 67700 & EVLA & $560\pm70$ & \\
\enddata
\end{deluxetable}

\begin{deluxetable}{ccccccccc}
\tabletypesize{\normalsize}
\tablecaption{Astrometry for \w1814\ 
\label{table:astrometry}}
\tablewidth{0pt}
\tablehead{
\colhead{Band} & \colhead{RA (J2000.0)} & \colhead{Decl. (J2000.0)} & \colhead{RA B} &
\colhead{Decl. B} & \colhead{RA C} & \colhead{Decl. C} &  \colhead{RA D} & \colhead{Decl. D}}
\startdata
{\it g'} & 18 14 17.32$\pm0.02$ & 34 12 24.9$\pm0.2$ & 17.34 & 27.6 & 17.28 & 30.3 & 17.13 & 21.9 \\
{\it r'} & 18 14 17.31$\pm0.02$ & 34 12 25.1$\pm0.2$ & 17.34 & 27.6 & 17.27 & 30.4 & 17.14 & 21.8 \\
{\it J} & 18 14 17.29$\pm0.02$ & 34 12 25.2$\pm0.2$ & 17.35 & 27.6 & 17.27 & 30.5 \\
$K_s$ & 18 14 17.27$\pm0.02$ & 34 12 25.3$\pm0.2$ & 17.34 & 27.7 & 17.27 & 30.5 & 17.12 & 22.0 \\
$[3.6]$ & 18 14 17.29$\pm0.02$ & 34 12 25.1$\pm0.2$ & 17.35 & 27.6 & 17.28 & 30.2 & 17.13 & 21.9 \\
$[4.5]$ & 18 14 17.29$\pm0.02$ & 34 12 25.1$\pm0.2$ & 17.33 & 27.7 & 17.29 & 30.2 & 17.13 & 22.0 \\
WISE & 18 14 17.30$\pm0.02$ & 34 12 25.0$\pm0.3$ \\
$350\mu$m & 18 14 17.31$\pm0.20$ & 34 12 25.5$\pm2.0$ \\
C-band high & 18 14 17.26$\pm0.05$ & 34 12 24.2$\pm1.2$ \\
C-band low & 18 14 17.33$\pm0.03$ & 34 12 24.9$\pm0.5$ \\
\enddata
\end{deluxetable}

\normalsize

\clearpage

\end{document}